\documentclass[a4paper,11pt]{article}
\pdfoutput=1 
\usepackage[T1]{fontenc} 

\makeatletter
\gdef\@fpheader{ }
\gdef\@journal{ }
\makeatother
\RequirePackage{amsmath}
\RequirePackage{amssymb}
\RequirePackage{epsfig}
\RequirePackage{graphicx}
\RequirePackage[numbers,sort&compress]{natbib}
\RequirePackage{color}
\RequirePackage[colorlinks=true
,urlcolor=blue
,anchorcolor=blue
,citecolor=blue
,filecolor=blue
,linkcolor=blue
,menucolor=blue
,pagecolor=blue
,linktocpage=true
,pdfproducer=medialab
,pdfa=true
]{hyperref}

\newif\ifnotoc\notocfalse
\newif\ifemailadd\emailaddfalse
\newif\iftoccontinuous\toccontinuousfalse
\makeatletter
\def\@subheader{\@empty}
\def\@keywords{\@empty}
\def\@abstract{\@empty}
\def\@xtum{\@empty}
\def\@dedicated{\@empty}
\def\@arxivnumber{\@empty}
\def\@collaboration{\@empty}
\def\@collaborationImg{\@empty}
\def\@proceeding{\@empty}
\def\@preprint{\@empty}

\newcommand{\subheader}[1]{\gdef\@subheader{#1}}
\newcommand{\keywords}[1]{\if!\@keywords!\gdef\@keywords{#1}\else%
\PackageWarningNoLine{\jname}{Keywords already defined.\MessageBreak Ignoring last definition.}\fi}
\renewcommand{\abstract}[1]{\gdef\@abstract{#1}}
\newcommand{\dedicated}[1]{\gdef\@dedicated{#1}}
\newcommand{\arxivnumber}[1]{\gdef\@arxivnumber{#1}}
\newcommand{\proceeding}[1]{\gdef\@proceeding{#1}}
\newcommand{\xtumfont}[1]{\textsc{#1}}
\newcommand{\correctionref}[3]{\gdef\@xtum{\xtumfont{#1} \href{#2}{#3}}}
\newcommand\jname{JHEP}
\newcommand\acknowledgments{\section*{Acknowledgments}}

\newcommand\preprint[1]{\gdef\@preprint{\hfill #1}}

\makeatother

\newcommand\note[2][]{%
\if!#1!%
\stepcounter{footnote}\footnotetext{#2}%
\else%
{\renewcommand\thefootnote{#1}%
\footnotetext{#2}}%
\fi}

\makeatletter
\newtoks\auth@toks
\renewcommand{\author}[2][]{%
  \if!#1!%
    \auth@toks=\expandafter{\the\auth@toks#2\ }%
  \else
    \auth@toks=\expandafter{\the\auth@toks#2$^{#1}$\ }%
  \fi
}
\makeatother
\makeatletter
\newtoks\affil@toks\newif\ifaffil\affilfalse
\newcommand{\affiliation}[2][]{%
\affiltrue
  \if!#1!%
    \affil@toks=\expandafter{\the\affil@toks{\item[]#2}}%
  \else
    \affil@toks=\expandafter{\the\affil@toks{\item[$^{#1}$]#2}}%
  \fi
}
\makeatother
\makeatletter
\newtoks\email@toks\newcounter{email@counter}%
\newcommand{\emailAdd}[1]{%
\emailaddtrue%
\ifnum\theemail@counter>0\email@toks=\expandafter{\the\email@toks, \@email{#1}}%
\else\email@toks=\expandafter{\the\email@toks\@email{#1}}%
\fi\stepcounter{email@counter}}
\newcommand{\@email}[1]{\href{mailto:#1}{\tt #1}}
\makeatother

\makeatletter
\newcommand*\collaboration[1]{\gdef\@collaboration{#1}}
\newcommand*\collaborationImg[2][]{\gdef\@collaborationImg{#2}}
\makeatletter
\newcommand\afterLogoSpace{\smallskip}
\newcommand\afterSubheaderSpace{\vskip3pt plus 2pt minus 1pt}
\newcommand\afterProceedingsSpace{\vskip21pt plus0.4fil minus15pt}
\newcommand\afterTitleSpace{\vskip23pt plus0.06fil minus13pt}
\newcommand\afterRuleSpace{\vskip23pt plus0.06fil minus13pt}
\newcommand\afterCollaborationSpace{\vskip3pt plus 2pt minus 1pt}
\newcommand\afterCollaborationImgSpace{\vskip3pt plus 2pt minus 1pt}
\newcommand\afterAuthorSpace{\vskip5pt plus4pt minus4pt}
\newcommand\afterAffiliationSpace{\vskip3pt plus3pt}
\newcommand\afterEmailSpace{\vskip16pt plus9pt minus10pt\filbreak}
\newcommand\afterXtumSpace{\par\bigskip}
\newcommand\afterAbstractSpace{\vskip16pt plus9pt minus13pt}
\newcommand\afterKeywordsSpace{\vskip16pt plus9pt minus13pt}
\newcommand\afterArxivSpace{\vskip3pt plus0.01fil minus10pt}
\newcommand\afterDedicatedSpace{\vskip0pt plus0.01fil}
\newcommand\afterTocSpace{\bigskip\medskip}
\newcommand\afterTocRuleSpace{\bigskip\bigskip}
\newlength{\affiliationsSep}
\newcommand\beforetochook{\pagestyle{myplain}\pagenumbering{roman}}

\DeclareFixedFont\trfont{OT1}{phv}{b}{sc}{11}

\renewcommand\maketitle{
\pagestyle{empty}
\thispagestyle{titlepage}
\setcounter{page}{0}
\noindent{\small\scshape\@fpheader}\@preprint\par

\afterLogoSpace
\if!\@subheader!\else\noindent{\trfont{\@subheader}}\fi
\afterSubheaderSpace
\if!\@proceeding!\else\noindent{\sc\@proceeding}\fi
\afterProceedingsSpace
{\LARGE\flushleft\sffamily\bfseries\@title\par}
\afterTitleSpace
\hrule height 1.5\p@%
\afterRuleSpace
\if!\@collaboration!\else
{\Large\bfseries\sffamily\raggedright\@collaboration}\par
\afterCollaborationSpace
\fi
\if!\@collaborationImg!\else
{\normalsize\bfseries\sffamily\raggedright\@collaborationImg}\par
\afterCollaborationImgSpace
\fi
{\bfseries\raggedright\sffamily\the\auth@toks\par}
\afterAuthorSpace
\ifaffil\begin{list}{}{%
\setlength{\leftmargin}{0.28cm}%
\setlength{\labelsep}{0pt}%
\setlength{\itemsep}{\affiliationsSep}%
\setlength{\topsep}{-\parskip}}
\itshape\small%
\the\affil@toks
\end{list}\fi
\afterAffiliationSpace
\ifemailadd 
\noindent\hspace{0.28cm}\begin{minipage}[l]{.9\textwidth}
\begin{flushleft}
\textit{E-mail:} \the\email@toks
\end{flushleft}
\end{minipage}
\else 
\PackageWarningNoLine{\jname}{E-mails are missing.\MessageBreak Plese use \protect\emailAdd\space macro to provide e-mails.}
\fi
\afterEmailSpace
\if!\@xtum!\else\noindent{\@xtum}\afterXtumSpace\fi
\if!\@abstract!\else\noindent{\renewcommand\baselinestretch{.9}\textsc{Abstract:}}\ \@abstract\afterAbstractSpace\fi
\if!\@keywords!\else\noindent{\textsc{Keywords:}} \@keywords\afterKeywordsSpace\fi
\if!\@arxivnumber!\else\noindent{\textsc{ArXiv ePrint:}} \href{http://arxiv.org/abs/\@arxivnumber}{\@arxivnumber}\afterArxivSpace\fi
\if!\@dedicated!\else\vbox{\small\it\raggedleft\@dedicated}\afterDedicatedSpace\fi
\ifnotoc\else
\iftoccontinuous\else\newpage\fi
\beforetochook\hrule
\tableofcontents
\afterTocSpace
\hrule
\afterTocRuleSpace
\fi
\setcounter{footnote}{0}
\pagestyle{myplain}\pagenumbering{arabic}
} 

\renewcommand{\baselinestretch}{1.1}\normalsize
\divide\@tempdima\baselineskip
\@tempcnta=\@tempdima
\skip\@mpfootins = \skip\footins
\renewcommand{\@dotsep}{10000}

\newcommand\ps@myplain{
\pagenumbering{arabic}
\renewcommand\@oddfoot{\hfill-- \thepage\ --\hfill}
\renewcommand\@oddhead{}}
\let\ps@plain=\ps@myplain

\newcommand\ps@titlepage{\renewcommand\@oddfoot{}\renewcommand\@oddhead{}}

\numberwithin{equation}{section}

\renewcommand\section{\@startsection{section}{1}{\z@}%
                                   {-3.5ex \@plus -1.3ex \@minus -.7ex}%
                                   {2.3ex \@plus.4ex \@minus .4ex}%
                                   {\normalfont\large\bfseries}}
\renewcommand\subsection{\@startsection{subsection}{2}{\z@}%
                                   {-2.3ex\@plus -1ex \@minus -.5ex}%
                                   {1.2ex \@plus .3ex \@minus .3ex}%
                                   {\normalfont\normalsize\bfseries}}
\renewcommand\subsubsection{\@startsection{subsubsection}{3}{\z@}%
                                   {-2.3ex\@plus -1ex \@minus -.5ex}%
                                   {1ex \@plus .2ex \@minus .2ex}%
                                   {\normalfont\normalsize\bfseries}}
\renewcommand\paragraph{\@startsection{paragraph}{4}{\z@}%
                                   {1.75ex \@plus1ex \@minus.2ex}%
                                   {-1em}%
                                   {\normalfont\normalsize\bfseries}}
\renewcommand\subparagraph{\@startsection{subparagraph}{5}{\parindent}%
                                   {1.75ex \@plus1ex \@minus .2ex}%
                                   {-1em}%
                                   {\normalfont\normalsize\bfseries}}

\def\fnum@figure{\textbf{\figurename\nobreakspace\thefigure}}
\def\fnum@table{\textbf{\tablename\nobreakspace\thetable}}

\long\def\@makecaption#1#2{%
  \vskip\abovecaptionskip
  \sbox\@tempboxa{\small #1. #2}%
  \ifdim \wd\@tempboxa >\hsize
    \small #1. #2\par
  \else
    \global \@minipagefalse
    \hb@xt@\hsize{\hfil\box\@tempboxa\hfil}%
  \fi
  \vskip\belowcaptionskip}

\renewenvironment{thebibliography}[1]{%
\begin{oldthebibliography}{#1}%
\small%
\raggedright%
\setlength{\itemsep}{5pt plus 0.2ex minus 0.05ex}%
}%
{%
\end{oldthebibliography}%
}

\begin{document}


\title{\boldmath Exact solution of inverse-square-root potential $V\left(  r\right)
=-\frac{\alpha}{\sqrt{r}}$}

\author[a]{Wen-Du Li}
\author[a,b,2]{and Wu-Sheng Dai}\note{daiwusheng@tju.edu.cn.}


\affiliation[a]{Department of Physics, Tianjin University, Tianjin 300072, P.R. China}
\affiliation[b]{LiuHui Center for Applied Mathematics, Nankai University \& Tianjin University, Tianjin 300072, P.R. China}









\abstract{An exact solution of the three-dimensional spherical symmetry
inverse-square-root potential $V\left(  r\right)  =-\frac{\alpha}{\sqrt{r}}$,
including scattering and bound-state solutions, is presented.
}

\maketitle
\flushbottom


\section{Introduction}

The inverse-square-root potential $V\left(  r\right)  =-\alpha/\sqrt{r}$ is a
long-range potential which has both scattering states and bound states. As a
long-range potential, the inverse-square-root potential has a much longer
range than the Coulomb potential. Its scattering state has a very different
asymptotic behavior from that of the Coulomb potential. Its bound state, other
than that of the Coulomb potential, has no closed classical orbits. An exact
solution of the inverse-square-root potential enables us to investigate the
behavior of long-range potentials deeply.

The inverse-power potential $V\left(  r\right)  \sim1/r^{s}$ with $0<s<2$ is a
long-range potential and has both scattering states and bound states; while
$V\left(  r\right)  \sim1/r^{s}$ with $s\geq2$ is a short-range potential and
has only scattering states. The short-range power potential, $1/r^{s}$ with
$s\geq2$, has only scattering states and can be generally treated
\cite{liu2014scattering,joachain1975quantum}. The long-range power potential,
$1/r^{s}$ with $0<s<2$, only when $s=1$, the Coulomb potential, is exactly
solved. In this paper, we present an exact solution of the inverse-square-root
potential ----- one other long-range potential.

The first thing we need to do is to determine the boundary condition, the
value of the wave function on the boundary. At infinity, for scattering
states, the wave function must equal the large-distance asymptotics; for bound
states, the wave function must equal zero.

Different long-range potentials have different scattering boundary conditions,
because the boundary condition of long-range-potential scattering at infinity
is the large-distance asymptotic solution of the potential and, in general,
the asymptotic solutions of different long-range potentials are different.
Therefore, we have to determine the scattering boundary condition one by one
for various long-range potentials, rather than short-range-potential
scattering in which all potentials have the same scattering boundary
condition. For this reason, in order to solve the inverse-square-root
potential, a long-range potential, we need to solve its large-distance
asymptotic solution first.

In section \ref{RE}, we convert the radial equation of the inverse-square-root
potential to the biconfluent Heun equation. In section \ref{RegularSolution},
we solve the regular solution and in section \ref{IrregularSolution} we solve
the irregular solution. The exact solutions of bound and scattering states are
given in section \ref{ExactSolution}. The conclusion is given in section
\ref{Conclusion}.

\section{Radial equation \label{RE}}

The radial wave function $R_{l}\left(  r\right)  =u_{l}\left(  r\right)  /r$
of the inverse-square-root potential
\begin{equation}
V\left(  r\right)  =-\frac{\alpha}{\sqrt{r}}%
\end{equation}
is jointly determined by the radial equation
\begin{equation}
\frac{d^{2}u_{l}\left(  r\right)  }{dr^{2}}+\left[  k^{2}-\frac{l\left(
l+1\right)  }{r^{2}}+\frac{\alpha}{\sqrt{r}}\right]  u_{l}\left(  r\right)
=0\label{jxfc}%
\end{equation}
and the boundary conditions at $r=0$ and at $r\rightarrow\infty$.

At $r=0$, both for scattering states and bound states, the boundary condition
is $u_{l}\left(  0\right)  =0$ (in fact, for the regular solution we use a
stronger condition $\lim_{r\rightarrow0}u_{l}\left(  r\right)  /r^{l+1}=1$).
At $r\rightarrow\infty$, for scattering states, the boundary condition is
$u_{l}\left(  r\rightarrow\infty\right)  =u_{l}^{\infty}\left(  r\right)  $,
where $u_{l}^{\infty}\left(  r\right)  $ is the large-distance asymptotic
solution of the radial equation (\ref{jxfc}); for bound states, the boundary
condition is $u_{l}\left(  r\rightarrow\infty\right)  =0$.

By introducing $z=\sqrt{-2ikr}$ and%
\begin{equation}
u_{l}\left(  z\right)  =A_{l}\exp\left(  -\left(  \frac{z^{2}}{2}+\lambda
z\right)  \right)  z^{2\left(  l+1\right)  }f_{l}\left(  z\right)  ,
\label{uf12}%
\end{equation}
where $\lambda=\alpha/\sqrt{2ik^{3}}$ and $A_{l}$ a constant, we convert the
radial equation (\ref{jxfc}) into an equation of $f_{l}\left(  z\right)  $:%
\begin{equation}
zf_{l}^{\prime\prime}\left(  z\right)  -\left[  2{z}^{2}+2\lambda z-\left(
4l+3\right)  \right]  f_{l}^{\prime}\left(  z\right)  +\left\{  \left[
\lambda^{2}-\left(  4l+4\right)  \right]  z-\left(  4l+3\right)
\lambda\right\}  f_{l}\left(  z\right)  =0. \label{eqfsqrt12}%
\end{equation}

The equation of $f_{l}\left(  z\right)  $, eq. (\ref{eqfsqrt12}), is just the
so-called biconfluent Heun equation \cite{ronveaux1995heun}.

\section{Regular solution \label{RegularSolution}}

The regular solution is a solution satisfying the boundary condition at $r=0$.
At $r=0$, the boundary condition for both bound states and scattering states
is \cite{romo1998study}
\begin{equation}
\lim_{r\rightarrow0}\frac{u_{l}\left(  r\right)  }{r^{l+1}}=1, \label{rsbc}%
\end{equation}
because the asymptotic solution of the radial equation (\ref{jxfc}) at $r=0$
is $u_{l}\left(  r\right)  \overset{r\rightarrow0}{\sim}r^{l+1}$
\cite{ballentine1998quantum}.

The biconfluent Heun equation, eq. (\ref{eqfsqrt12}), has two linearly
independent solutions \cite{ronveaux1995heun},%
\begin{align}
y_{l}^{\left(  1\right)  }\left(  z\right)   &  =N\left(  4l+2,2\lambda
,\lambda^{2},0,z\right)  ,\\
y_{l}^{\left(  2\right)  }\left(  z\right)   &  =cN\left(  4l+2,2\lambda
,\lambda^{2},0,z\right)  \ln z+\sum_{n\geq0}d_{n}z^{n},
\end{align}
where $N\left(  \alpha,\beta,\gamma,\delta,z\right)  $ is the Heun biconfluent
function \cite{ronveaux1995heun,Slavyanov2000special}, the constant%

\begin{equation}
c=\frac{1}{4l+2}\left[  d_{4l+1}\lambda\left(  4l+1\right)  -d_{4l}\left(
\lambda^{2}-4l\right)  \right]  ,
\end{equation}
and the coefficient $d_{\nu}$ is given by the recurrence relation%
\begin{align}
&  d_{-1}=0,\text{ \ }d_{0}=1,\\
&  \left(  \nu+2\right)  \left(  \nu-4l\right)  d_{\nu+2}-\lambda\left(
2\nu+1-4l\right)  d_{\nu+1}+\left[  \lambda^{2}-2\left(  \nu+1\right)
+4l+2\right]  d_{\nu}=0.
\end{align}

The solution can be determined by the boundary condition (\ref{rsbc}).

The Heun biconfluent function has the expansion \cite{ronveaux1995heun}%
\begin{equation}
N\left(  4l+2,2\lambda,\lambda^{2},0,z\right)  =\sum_{n\geq0}\frac{A_{n}%
}{\left(  4l+3\right)  _{n}}\frac{z^{n}}{n!},
\end{equation}
where the coefficient $A_{n}$ is given by%
\begin{align}
A_{0}  &  =1,\text{ \ }A_{1}=\left(  4l+3\right)  \lambda,\\
A_{n+2}  &  =\lambda\left(  4l+2n+5\right)  A_{n+1}-\left(  n+1\right)
\left(  4l+n+3\right)  \left[  \lambda^{2}-\left(  4l+2n+4\right)  \right]
A_{n},
\end{align}
and $\left(  a\right)  _{n}=\Gamma\left(  a+n\right)  /\Gamma\left(  a\right)
$ is Pochhammer's symbol.

Obviously, only $f_{l}\left(  z\right)  =y_{l}^{\left(  1\right)  }\left(
z\right)  $ satisfies the boundary condition of the regular solution, eq.
(\ref{rsbc}). By eq. (\ref{uf12}) and $z=\sqrt{-2ikr}$, we arrive at%
\begin{equation}
u_{l}\left(  r\right)  =A_{l}\left(  -2ikr\right)  ^{l+1}\exp\left(  i\left(
kr+\frac{\alpha}{k}\sqrt{r}\right)  \right)  N\left(  4l+2,2\lambda
,\lambda^{2},0,\sqrt{-2ikr}\right)  . \label{regular}%
\end{equation}
This is the regular solution.

\section{Irregular solution \label{IrregularSolution}}

The irregular solution is a solution satisfying the boundary condition at
$r\rightarrow\infty$. At $r\rightarrow\infty$, the bound state and the
scattering state require different boundary conditions.

\subsection{Scattering boundary condition}

The scattering boundary condition is determined by the asymptotic solution of
the radial equation (\ref{jxfc}) at $r\rightarrow\infty$. For long-range
potentials, the asymptotic behavior is determined by the potential, so
different potentials give different scattering boundary conditions. As a
comparison, for short-range potentials, the large-distance asymptotic behavior
is determined by the centrifugal potential $l\left(  l+1\right)  /r^{2}$ which
indeed comes from the kinetic energy rather than the external potential, so
the asymptotic solution and, then, the scattering boundary condition, are the
same for all short-range potentials.

To determine the scattering boundary condition, we first solve the asymptotic
solution of the radial equation (\ref{jxfc}).

Writing the radial wave function $u_{l}\left(  r\right)  $ as%
\begin{equation}
u_{l}\left(  r\right)  =e^{h\left(  r\right)  }\exp\left(  \pm ikr\right)
\label{uh}%
\end{equation}
and substituting into the radial equation (\ref{jxfc}) give the equation of
$h\left(  r\right)  $:
\begin{equation}
h^{\prime\prime}\left(  r\right)  +\left[  h^{\prime}\left(  r\right)
\right]  ^{2}\pm2ikh^{\prime}\left(  r\right)  =\frac{l\left(  l+1\right)
}{r^{2}}-\frac{\alpha}{\sqrt{r}}. \label{eqh}%
\end{equation}
The leading-order contributions on the right-hand side and the left-hand side
of eq. (\ref{eqh}) must be the same order of magnitude, so we have%
\begin{equation}
\pm2ikh^{\prime}\left(  r\right)  \overset{r\rightarrow\infty}{\sim}%
-\frac{\alpha}{\sqrt{r}}.
\end{equation}
This gives
\begin{equation}
h\left(  r\right)  \sim\pm i\frac{\alpha}{k}\sqrt{r}. \label{hlr}%
\end{equation}

Repeating the procedure in eq. (\ref{uh}) and taking the result given by eq.
(\ref{hlr}) into consideration, we write the radial wave function
$u_{l}\left(  r\right)  $ as
\begin{equation}
u_{l}\left(  r\right)  =e^{g\left(  r\right)  }\exp\left(  \pm i\left(
kr+\frac{\alpha}{k}\sqrt{r}\right)  \right)
\end{equation}
and substitute into the radial equation (\ref{jxfc}). We then have the
equation of $g\left(  r\right)  $:
\begin{equation}
g^{\prime\prime}\left(  r\right)  +\left[  g^{\prime}\left(  r\right)
\right]  ^{2}\pm2ikg^{\prime}\left(  r\right)  \pm\frac{i\alpha}{k\sqrt{r}%
}g^{\prime}\left(  r\right)  =\frac{l\left(  l+1\right)  }{r^{2}}+\frac
{\alpha^{2}}{4k^{2}r}\pm\frac{i\alpha}{4kr^{3/2}}.
\end{equation}
By the same reason, we have%
\begin{equation}
\pm2ikg^{\prime}\left(  r\right)  \overset{r\rightarrow\infty}{\sim}%
\frac{\alpha^{2}}{4k^{2}r}.
\end{equation}
This gives
\begin{equation}
g\left(  r\right)  \sim\mp i\frac{\alpha^{2}}{8k^{3}}\ln r.
\end{equation}

Again, repeating the above procedure, we write the radial wave function
$u_{l}\left(  r\right)  $ as
\begin{equation}
u_{l}\left(  r\right)  =e^{v\left(  r\right)  }\exp\left(  \pm i\left(
kr+\frac{\alpha}{k}\sqrt{r}-\frac{\alpha^{2}}{8k^{3}}\ln r\right)  \right)
\end{equation}
and substitute into the radial equation (\ref{jxfc}). We then have the
equation of $v\left(  r\right)  $:%
\begin{equation}
v^{\prime\prime}\left(  r\right)  +\left[  v^{\prime}\left(  r\right)
\right]  ^{2}\pm2ikv^{\prime}\left(  r\right)  \pm\frac{i\alpha}{k\sqrt{r}%
}v^{\prime}\left(  r\right)  \mp\frac{i\alpha^{2}}{4k^{3}r}v^{\prime}\left(
r\right)  =\frac{l\left(  l+1\right)  }{r^{2}}+\frac{\alpha^{4}}{64k^{6}r^{2}%
}\mp\frac{i\alpha^{2}}{8k^{3}r^{2}}-\frac{\alpha^{3}}{8k^{4}r^{3/2}}\pm
\frac{i\alpha}{4kr^{3/2}}.
\end{equation}
By the same reason, we have%
\begin{equation}
\pm2ikv^{\prime}\left(  r\right)  =-\frac{\alpha^{3}}{8k^{4}r^{3/2}}\pm
\frac{i\alpha}{4kr^{3/2}}.
\end{equation}
It can be directly seen that%
\begin{equation}
v\left(  r\right)  \sim\frac{1}{\sqrt{r}}.
\end{equation}
This means that when $r\rightarrow\infty$, the contribution of $v\left(
r\right)  $ vanishes and does not need to be taken into account.

We can now write down the asymptotic radial function:%
\begin{equation}
u_{l}^{\infty}\left(  r\right)  =\exp\left(  \pm i\left(  kr+\frac{\alpha}%
{k}\sqrt{r}-\frac{\alpha^{2}}{8k^{3}}\ln r\right)  \right)  . \label{sbc}%
\end{equation}
The scattering boundary condition, then, can be written as%
\begin{equation}
\lim_{r\rightarrow\infty}\exp\left(  \pm i\left(  kr+\frac{\alpha}{k}\sqrt
{r}-\frac{\alpha^{2}}{8k^{3}}\ln r\right)  \right)  u_{l}\left(  r\right)  =1.
\label{SC}%
\end{equation}
It is worth to compare the scattering boundary condition (\ref{SC}) with the
scattering boundary condition of the Coulomb potential, $\lim_{r\rightarrow
\infty}\exp\left(  \pm i\left(  kr-\frac{\alpha}{2k}\ln r\right)  \right)
u_{l}\left(  kr\right)  =1$ \cite{romo1998study}.

\subsection{Irregular solution}

With the scattering boundary condition at $r\rightarrow\infty$, eq.
(\ref{SC}), we can now determine the irregular solution which is the solution
satisfying the boundary condition at $r\rightarrow\infty$.

The biconfluent Heun equation (\ref{eqfsqrt12}), which relates the radial
equation by the substitution (\ref{uf12}), has two linearly independent
solutions satisfying the scattering boundary condition (\ref{SC})
\cite{ronveaux1995heun}%

\begin{align}
B_{l}^{+}\left(  4l+2,2\lambda,\lambda^{2},0,z\right)   &  =z^{\left(
\lambda^{2}-4l-4\right)  /2}\sum_{n\geq0}\frac{a_{n}}{z^{n}},\label{f112}\\
H_{l}^{+}\left(  4l+2,2\lambda,\lambda^{2},0,z\right)   &  =z^{-\left(
\lambda^{2}+4l+4\right)  /2}e^{2\lambda z+z^{2}}\sum_{n\geq0}\frac{e_{n}%
}{z^{n}}, \label{f212}%
\end{align}
where $B_{l}^{+}\left(  \alpha,\beta,\gamma,\delta,z\right)  $ and $H_{l}%
^{+}\left(  \alpha,\beta,\gamma,\delta,z\right)  $ are another two kinds of
the biconfluent Heun functions, which is different from the biconfluent Heun
functions mentioned above, and the coefficients $a_{n}$ and $e_{n}$ are given
by
\begin{align}
&  a_{0}=1,\text{ }a_{1}=-\frac{1}{2}\lambda\left(  \lambda^{2}+1\right)  ,\\
&  2\left(  n+2\right)  a_{n+2}+\left[  \lambda^{3}+\left(  2n+3\right)
\lambda\right]  a_{n+1}+\left[  \frac{1}{4}\lambda^{4}-\left(  n+1\right)
\lambda^{2}+n\left(  n+2\right)  -4l\left(  l+1\right)  \right]  a_{n}=0
\end{align}
and
\begin{align}
&  e_{0}=1,\text{ }e_{1}=\frac{1}{2}\lambda\left(  \lambda^{2}-1\right)  ,\\
&  2\left(  n+2\right)  e_{n+2}-\left[  \lambda^{3}-\left(  2n+3\right)
\lambda\right]  e_{n+1}-\left[  \frac{1}{4}\lambda^{4}+\left(  n+1\right)
\lambda^{2}+n\left(  n+2\right)  -4l\left(  l+1\right)  \right]  e_{n}=0.
\end{align}
These two solutions both are irregular solutions.

\section{Bound state and scattering state \label{ExactSolution}}

In order to achieve the bound-state and scattering-state solutions, we first
express the regular solution (\ref{regular}) as a linear combination of the
two irregular solutions \cite{ronveaux1995heun}:%

\begin{align}
N\left(  4l+2,2\lambda,\lambda^{2},0,\sqrt{-2ikr}\right)   &  =K_{1}\left(
4l+2,2\lambda,\lambda^{2},0\right)  B_{l}^{+}\left(  4l+2,2\lambda,\lambda
^{2},0,\sqrt{-2ikr}\right) \nonumber\\
&  +K_{2}\left(  4l+2,2\lambda,\lambda^{2},0\right)  H_{l}^{+}\left(
4l+2,2\lambda,\lambda^{2},0,\sqrt{-2ikr}\right)  , \label{BH}%
\end{align}
where $K_{1}\left(  4l+2,2\lambda,\lambda^{2},0\right)  $ and $K_{2}\left(
4l+2,2\lambda,\lambda^{2},0\right)  $ are the coefficients of combination.
Then by eq. (\ref{regular}) and eqs. (\ref{f112}) and (\ref{f212}), we have%
\begin{align}
u_{l}\left(  r\right)   &  =A_{l}K_{1}\left(  4l+2,2\lambda,\lambda
^{2},0\right)  \left(  -2ik\right)  ^{-i\alpha^{2}/\left(  8k^{3}\right)
}\exp\left(  i\left(  kr+\frac{\alpha}{k}\sqrt{r}-\frac{\alpha^{2}}{8k^{3}}\ln
r\right)  \right)  \sum_{n\geq0}\frac{a_{n}}{\left(  -2ikr\right)  ^{n/2}%
}\nonumber\\
&  +A_{l}K_{2}\left(  4l+2,2\lambda,\lambda^{2},0\right)  \left(  -2ik\right)
^{i\alpha^{2}/\left(  8k^{3}\right)  }\exp\left(  -i\left(  kr+\frac{\alpha
}{k}\sqrt{r}-\frac{\alpha^{2}}{8k^{3}}\ln r\right)  \right)  \sum_{n\geq
0}\frac{e_{n}}{\left(  -2ikr\right)  ^{n/2}}. \label{radialu}%
\end{align}
Here, we use the expansions (\ref{f112}) and (\ref{f212}) for convenience in
analyzing the asymptotic behavior of the solution.

\subsection{Bound state}

By analytical continuation $k$ to the complex plane, we can consider $k$ along
the positive imaginary axis. On the positive imaginary axis, we define%
\begin{equation}
k=i\kappa,\text{ \ }\kappa>0.
\end{equation}

Then eq. (\ref{radialu}), with $\lambda=\alpha/\sqrt{2\kappa^{3}}$, becomes%
\begin{align}
u_{l}\left(  r\right)   &  =A_{l}K_{1}\left(  4l+2,2\lambda,\lambda
^{2},0\right)  \exp\left(  -\kappa r+\frac{\alpha}{\kappa}\sqrt{r}\right)
\left(  2\kappa r\right)  ^{\alpha^{2}/\left(  8\kappa^{3}\right)  }%
\sum_{n\geq0}\frac{a_{n}}{\left(  2\kappa r\right)  ^{n/2}}\nonumber\\
&  +A_{l}K_{2}\left(  4l+2,2\lambda,\lambda^{2},0\right)  \exp\left(  \kappa
r-\frac{\alpha}{\kappa}\sqrt{r}\right)  \left(  2\kappa r\right)
^{-\alpha^{2}/\left(  8\kappa^{3}\right)  }\sum_{n\geq0}\frac{e_{n}}{\left(
2\kappa r\right)  ^{n/2}}.
\end{align}
\qquad\qquad\qquad

It can be directly seen that the first term vanishes when $r\rightarrow\infty$
due to the factor $\exp\left(  -\kappa r+\alpha\sqrt{r}/\kappa\right)  $ and,
in contrast, the second term diverges when $r\rightarrow\infty$ due to the
factor $\exp\left(  \kappa r-\alpha\sqrt{r}/\kappa\right)  $. Clearly, when
$K_{2}\left(  4l+2,2\lambda,\lambda^{2},0\right)  $ equals zero, the solution
$u_{l}\left(  r\right)  $ will satisfy the bound-state boundary condition.
That is to say, the zeroes of the coefficient $K_{2}\left(  4l+2,2\lambda
,\lambda^{2},0\right)  $ corresponds to the bound state. This implies that the
zeroes of $K_{2}\left(  4l+2,2\lambda,\lambda^{2},0\right)  $ on the imaginary
axis determines the spectrum of the bound state. Consequently, the spectrum of
the bound-state eigenvalue is determined by
\begin{equation}
K_{2}\left(  4l+2,\frac{2\alpha}{\sqrt{2\kappa^{3}}},\frac{\alpha^{2}}%
{2\kappa^{3}},0\right)  =0,\label{K2spectrum}%
\end{equation}
where \cite{ronveaux1995heun}%
\begin{equation}
K_{2}\left(  \alpha,\beta,\gamma,\delta\right)  =\frac{\Gamma\left(
1+\alpha\right)  }{\Gamma\left(  \left(  \alpha-\gamma\right)  /2\right)
\Gamma\left(  1+\left(  \alpha+\gamma\right)  /2\right)  }J_{1+\left(
\alpha+\gamma\right)  /2}\left(  \frac{1}{2}\left(  \alpha+\gamma\right)
,\beta,\frac{1}{2}\left(  3\alpha-\gamma\right)  ,\delta+\frac{1}{2}%
\beta\left(  \gamma-\alpha\right)  \right)  ,
\end{equation}
with%
\begin{equation}
J_{\lambda}\left(  \alpha,\beta,\gamma,\delta\right)  =\int_{0}^{\infty
}x^{\lambda-1}e^{-x^{2}-\beta x}N\left(  \alpha,\beta,\gamma,\delta,x\right)
dx.
\end{equation}

Eq. (\ref{K2spectrum}) is an implicit expression of the bound-state spectrum.

The bound-state eigenfunction is then%
\begin{equation}
u_{l}\left(  r\right)  =C\exp\left(  -\kappa r+\frac{\alpha}{\kappa}\sqrt
{r}\right)  \left(  2\kappa r\right)  ^{\alpha^{2}/\left(  8\kappa^{3}\right)
}\sum_{n\geq0}\frac{a_{n}}{\left(  2\kappa r\right)  ^{n/2}},
\end{equation}
where the eigenvalue $\kappa$ is given by eq. (\ref{K2spectrum}) and $C$ is a
normalization constant.

\subsection{Scattering state}

It is known that the singularity on the positive imaginary axis of the
$S$-matrix corresponds to the spectrum of bound states
\cite{joachain1975quantum}. While, as pointed above, the zeroes on the
positive imaginary of the coefficient $K_{2}\left(  4l+2,2\lambda,\lambda
^{2},0\right)  $ correspond to the spectrum of bound states. Therefore, the
zeroes on the positive imaginary of $K_{2}\left(  4l+2,2\lambda,\lambda
^{2},0\right)  $ are just the singularities of the $S$-matrix. Considering
that the $S$-matrix%
\begin{equation}
S_{l}=e^{2i\delta_{l}} \label{Sl}%
\end{equation}
is unitary, we then have%
\begin{equation}
S_{l}=\frac{K_{2}^{\ast}\left(  4l+2,2\lambda,\lambda^{2},0\right)  }%
{K_{2}\left(  4l+2,2\lambda,\lambda^{2},0\right)  }=\frac{K_{2}\left(
4l+2,-2i\lambda,-\lambda^{2},0\right)  }{K_{2}\left(  4l+2,2\lambda
,\lambda^{2},0\right)  }, \label{SMat}%
\end{equation}
where $K_{2}^{\ast}\left(  4l+2,\frac{2\alpha}{\sqrt{2ik^{3}}},\frac
{\alpha^{2}}{2ik^{3}},0\right)  =K_{2}\left(  4l+2,\frac{2\alpha}%
{\sqrt{-2ik^{3}}},-\frac{\alpha^{2}}{2ik^{3}},0\right)  $ is used.

By the $S$-matrix (\ref{SMat}), we can construct the scattering wave function.
The scattering wave function can be generally written as $u_{l}\left(
r\right)  =A_{l}\left[  \left(  -1\right)  ^{l+1}u_{in}+S_{l}u_{out}\right]  $
with $u_{in}=u_{out}^{\ast}$, where $u_{in}$ and $u_{out}$ are the radially
ingoing and outgoing waves, respectively \cite{joachain1975quantum}.

Now we can write down the scattering wave function%
\begin{align}
u_{l}\left(  r\right)   &  =A_{l}\left[  \left(  -1\right)  ^{l+1}\exp\left(
-i\left(  kr+\frac{\alpha}{k}\sqrt{r}-\frac{\alpha^{2}}{8k^{3}}\ln r\right)
\right)  \sum_{n\geq0}\frac{e_{n}}{\left(  -2ikr\right)  ^{n/2}}\right.
\nonumber\\
&  +\left.  \frac{K_{2}\left(  4l+2,2i\lambda,-\lambda^{2},0\right)  }%
{K_{2}\left(  4l+2,2\lambda,\lambda^{2},0\right)  }\exp\left(  i\left(
kr+\frac{\alpha}{k}\sqrt{r}-\frac{\alpha^{2}}{8k^{3}}\ln r\right)  \right)
\sum_{n\geq0}\frac{e_{n}^{\ast}}{\left(  2ikr\right)  ^{n/2}}\right]  .
\end{align}

At $r\rightarrow\infty$, we have the large-distance asymptotics%
\begin{align}
&  u_{l}\left(  r\right)  \overset{r\rightarrow\infty}{\sim}A_{l}i^{l}\left[
-i^{l}\exp\left(  -i\left(  kr+\frac{\alpha}{k}\sqrt{r}-\frac{\alpha^{2}%
}{8k^{3}}\ln r\right)  \right)  \right.  \nonumber\\
&  +\left.  i^{-l}\frac{K_{2}\left(  4l+2,-2i\lambda,-\lambda^{2},0\right)
}{K_{2}\left(  4l+2,2\lambda,\lambda^{2},0\right)  }\exp\left(  i\left(
kr+\frac{\alpha}{k}\sqrt{r}-\frac{\alpha^{2}}{8k^{3}}\ln r\right)  \right)
\right]  ,\nonumber\\
&  =A_{l}i^{l}e^{i\delta_{l}}\sin\left(  kr+\frac{\alpha}{k}\sqrt{r}%
-\frac{\alpha^{2}}{8k^{3}}\ln r+\delta_{l}-\frac{l\pi}{2}\right)
.\label{uscaasy}%
\end{align}
The scattering phase shift here, by eqs. (\ref{Sl}) and (\ref{SMat}), is%
\begin{equation}
\delta_{l}=-\arg K_{2}\left(  4l+2,2\lambda,\lambda^{2},0\right)  .
\end{equation}

\section{Conclusion \label{Conclusion}}

In this paper, we present an exact solution of the spherical symmetry
inverse-power potential $-\alpha/\sqrt{r}$.

It is worthy to emphasized here that we can put a stronger definition for
long-range potentials: if the asymptotic behavior of a potential is the same
as that of short-range potentials, i.e., large-distance asymptotics is $e^{\pm
ikr}$, the potential is a short-range potential. Under this definition, the
inverse-power potential $V\left(  r\right)  \sim1/r^{s}$ with $0<s<1$ is a
long-range potential, but $V\left(  r\right)  \sim1/r^{s}$ with $1<s<2$ is a
short-range potential. Even under such a stronger definition, the
inverse-power potential $\alpha/\sqrt{r}$ is still a long-range potential.

The three-dimensional inverse square root potential is a long-range potential.
The difficulty of the study of long-range-potential scattering is that the
scattering boundary condition is determined by the external potential rather
than the centrifugal potential as that in short-range-potential scattering.
Scattering boundary conditions for different large-distance potentials are
different. To seek the scattering boundary condition for long-range potentials
needs to first determine the large-distance asymptotic solution. There are
some discussions on the long-range-potential scattering
\cite{enss1979asymptotic,levy1963low,hinckelmann1971low}. Scattering by
combination of known long-range and unknown short-range potentials is studied
by the renormalization-group method \cite{barford2003renormalization}. The
late-time dynamics of the wave equation with a long-range potential is
discussed in ref. \cite{hod2013scattering}. The spectral property of
scattering matrix of the Schr\"{o}dinger operator with a long-range potential
is considered in ref. \cite{yafaev1998scattering}. Scattering on black holes
is essentially a kind of long-range scattering.
\cite{stadnik2013resonant,flambaum2012dense}. The partial derivative of
scattering phase shifts with respect to wave number $k$ for long-range
potentials is given in ref. \cite{romo1998study}. There are many efforts on
seeking the scattering phase shift \cite{pang2012relation,li2015heat}. There
are also studies on the orbit in power-law potentials
\cite{grant1994classical}, the screened Coulomb potential, and the isotropic
harmonic oscillator \cite{wu2000dynamical}. The number of bound states
\cite{bargmann1952number,calogero1965upper}, the number of eigenstates
\cite{dai2009number,dai2010approach}, the conditions for the existence of
bound states \ \cite{brau2003necessary,brau2004sufficient} are also important problems.


\acknowledgments

We are very indebted to Dr G. Zeitrauman for his encouragement. This work is supported in part by NSF of China under Grant
No. 11575125 and No. 11375128.










\providecommand{\href}[2]{#2}\begingroup\raggedright\endgroup


\end{document}